# TARGET LOCALIZATION IN WIRELESS SENSOR NETWORKS BASED ON RECEIVED SIGNAL STRENGTH

Santhosh N Bharadwaj, Jagadeesha S N and Ravindra S

Department of Computer Science and Engineering,
Jawaharlal Nehru National College of Engineering, Shimoga, India

*ABSTRACT*

*We consider the problem of localizing a target taking the help of a set of anchor beacon nodes. A small number of beacon nodes are deployed at known locations in the area. The target can detect a beacon provided it happens to lie within the beacon's transmission range. Thus, the target obtains a measurement vector containing the readings of the beacons: '1' corresponding to a beacon if it is able to detect the target, and '0' if the beacon is not able to detect the target. The goal is twofold: to determine the location of the target based on the binary measurement vector at the target; and to study the behaviour of the localization uncertainty as a function of the beacon transmission range (sensing radius) and the number of beacons deployed. Beacon transmission range means signal strength of the beacon to transmit and receive the signals which is called as Received Signal Strength (RSS). To localize the target, we propose a grid-mapping based approach, where the readings corresponding to locations on a grid overlaid on the region of interest are used to localize the target. To study the behaviour of the localization uncertainty as a function of the sensing radius and number of beacons, extensive simulations and numerical experiments are carried out. The results provide insights into the importance of optimally setting the sensing radius and the improvement obtainable with increasing number of beacons.*

*KEYWORDS*

*Target Localization, uncertainty, Grid mapping based approach, Received Signal Strength (RSS)*

## 1. INTRODUCTION

In wireless sensor networks, nodes are deployed into an unplanned infrastructure where there is no a priori knowledge of location. The main objective is to locate each node as accurately as possible with a certain amount of smaller uncertainty. To identify the coordinates of sensor nodes (also called unknown nodes) require measuring a distance e.g., measuring time of arrival (ToA) or time difference of arrival (TdoA). Difficulties concerning time measurement result from synchronization of involved devices as well as the high mathematical effort to calculate the position. The measurement of the received signal strength (RSS) offers a possibility to determine distance with minimal effort.

The estimation of the position of an unknown sensor node from a set of measurements is called Localization. Ranging methods aim at estimating the distance of a receiver to a transmitter, by

DOI : 10.5121/sipij.2015.6302   11

exploiting known signal characteristics. For example, pairs of nodes in a sensor network whose radii are in communication range – beacon transmission range, of each other can use Received Signal Strength (RSS) techniques to estimate the RF signal strength of the receiver.

Circular domain means sensors are deployed uniformly and arbitrary domain means sensors are randomly distributed [2]. One fundamental issue in sensor networks is the coverage problem, which reflects how target localization is done [4]. The stochastic coverage of sensor network model is used where stochastic random distribution model can be uniform, Gaussian or any other distribution based on application [9]. Intersection problem of sensors, overlap of sensors, randomization of sensors and field of interest [5] are to be considered while working in wireless sensor networks.

If the source signal strength is known, along with the attenuation law for signal strength as a function of distance, then the receiver node can use RSS to estimate its distance from the sender. Such a distance estimate, however, is usually not very accurate because RSS can vary substantially owing to fading, shadowing, and multi path effects. Variations in height between sender and receiver can also affect the measurement accuracy. Furthermore, optimum radii in typical sensor nodes, number of optimum sensors required should be selected properly, otherwise it may cause severe uncertainty, because of cost considerations. These sensors do not come with well- calibrated components [7], therefore the source signal strength value may exhibit significant fluctuations [6].

In this paper, to determine the least possible uncertainty, randomization of beacons [5] , is done in a bounded domain [2]. Starting from least number, varying the number of beacons and varying radii of the beacons the uncertainty is calculated. Then graph is plotted on beacon radius v/s localization uncertainty and number of beacons v/s localization uncertainty, which gives the optimum beacon radius and optimum number of beacons respectively required to achieve the least possible uncertainty in localization. Deviations in uncertainty followed by COV is also calculated.

The paper is organized as follows. Section 2 discusses problem formulation, system model and algorithm to find the localization uncertainty. In Section 3 simulations and results are discussed followed by conclusion in Section 4.

## 2. PROBLEM FORMULATION

Received signal strength (RSS) is a range based localization technique which is used to estimate the distance between two nodes based on the signal strength received by the another node. The two nodes can be a beacon or sensor and a target. Received signal strength is inversely proportional to square of the distance.

$$RSS \propto \frac{1}{(\text{distance})^2} \tag{1}$$

As the distance between the beacon and target increases, the received signal strength decreases. Basically RSS deals with the estimation of the signal strength, so the communication range or beacon transmission range decreases with increase in distance is highly affected to changes in environmental conditions [10] like temperature, pressure, humidity, rainfall, fog, mist, drastic variations in climate, natural calamities like earthquake, Tsunami, thunder storms and the signal strength also attenuates due to obstacles [11] and this results to reflections, diffraction, scattering, interference of signals, polarization of signals and many more. All these parameters cause the signal strength to decrease. So modelling RSS mathematically is very difficult due to the above observations.

The received signal strength or received power, $P_r$ is related to distance, $d$ and is given by

$$\frac{P_r}{P_t} = G_t . G_r \left(\frac{\lambda}{4\pi d}\right)^2 \quad (2)$$

where, $P_r$ is received power, $P_t$ is transmitted power, $G_t$ and $G_r$ are transmitter and receiver antenna gains respectively. $\lambda$ is wavelength of signal transmitted and $d$ is distance between antennas. This equation is called as Friis equation.
Equation (2) can be modified as,

$$P(d)[dBm] = P(d_0)[dBm] - 10\gamma \log_{10}\left(\frac{d}{d_0}\right) + X_\sigma \quad (3)$$

where $P(d_0)$ represents the transmitting power of a wireless device (beacon) at the reference distance $d_0$, i.e., beacon transmission range, $d$ is the distance between the wireless device (beacon) and the access point (target), $\gamma$ is the path loss exponent and $X_\sigma$ is the shadow fading which follows zero mean Gaussian distribution with σ standard deviation.

Uncertainty in wireless sensor networks should be modelled carefully or otherwise severe uncertainties will affects the system. Localization is effective in nature only when a target is located with least possible uncertainty. Strictly speaking uncertainty depends on transmission range of beacons (sensing range) or beacon radius and number of beacons deployed in a bounded domain. Target location is found based on binary measurement vector and behaviour of localization uncertainty is studied as a function of beacon transmission range and number of beacons deployed.

The function of uncertainty in localization, f (r , b) is calculated by using two parameters: one is beacon radius, r and other one is number of beacons, b. Then the localization uncertainty is judged by means of increasing number of beacons and radius of the beacons. Then graph is plotted with varying number of beacons v/s uncertainty in localization and varying beacon radius v/s uncertainty in localization to have a clear picture about the least possible uncertainty that can be obtained. This analysis gives the optimum beacon radius and optimum number of beacons required to achieve the least possible localization uncertainty. Standard deviation in uncertainty is calculated to know the variations in localization uncertainty. And Co-efficient Of Variation (COV) is also calculated. COV is a standard measure of randomization of beacons. It verifies whether the number of beacons and also beacon radius calculated are optimum in nature and checks about the least possible uncertainty that is achieved and also possible variations in uncertainty.

## 2.1. System model

Let $S_1, S_2, S_3, ..., S_n$ be beacons and $r_1, r_2, r_3, ......, r_n$ be the normalized radii of the beacons. To identify the vicinity of the target (point $x$), which is deployed in an unknown location from known beacons positions in a bounded region [2], the condition is $\|x - s_i\|^2 < r_i^2$. This condition gives the distance from beacons to target (point $x$). This distance is known as Euclidean distance. This Euclidean distance returns the $n$ - dimensional binary vector, $U_i$. Select a fine grid of points and find readings of all points. Find points in the fine grid with reading, $U_i$, calculate area of points with same reading and area of uncertainty of targets is calculated by dividing the area of points with the same reading to the total area. This gives the percentage of uncertainty in localization.

## 2.2. Algorithm development

The algorithm can be designed as follows:

1. Given a point $x$, find its reading. Let $S_1, S_2, S_3,..., S_n$, are beacons, $r_1, r_2, r_3,....., r_n$ are radii of the beacons [4].

2. To identify the vicinity of the target [3], which is deployed in a unknown location from known beacons positions in a bounded region, the condition is

$$\|x - s_i\|^2 < r_i^2 \qquad (4)$$

   this condition gives the distance from beacons to target (point $x$). This distance is known as Euclidean distance.

3. This Euclidean distance also returns the $n$ - dimensional binary vector [8],[12],[13] reading, $U_i$ with a condition

$$\|x - s_i\|^2 < r_i^2 \quad \begin{cases} return & U_i = 1 \\ else & U_i = 0 \end{cases}$$

4. Select a fine grid of points and find readings of all points. Find points in the fine grid with reading, $U_i$. Grid mapping based approach is used here.

5. Localization uncertainty and binary measurement vector is found by using grid mapping based approach. Here the entire bounded region or area considered will be integrated to unit squares area. For instance, let the area = 100 * 100, a square with normalized units. Integrate this area by unit square area (1 * 1 = 1 sq. units), so that total number of squares in area 100 * 100 is 10000 squares = area of 10000 sq. units. This is known as Grid mapping based approach and is shown in Figure 1.

6. Calculate area of points with same reading and area of uncertainty of target [1], from the following expression:

$$\frac{area}{Total\ area} * 100 = \text{Percentage of Uncertainty} \qquad (5)$$

7. This percentage of uncertainty i.e., localization uncertainty that is obtained should be the least possible one, which is analyzed by means of graph.

8. Beacon radius v/s localization uncertainty gives the optimum beacon radius required to achieve least possible uncertainty.

9. Number of beacons v/s localization uncertainty gives the optimum number of beacons required to achieve least possible uncertainty.

10. Calculate Standard deviation and Co – efficient of Variation.

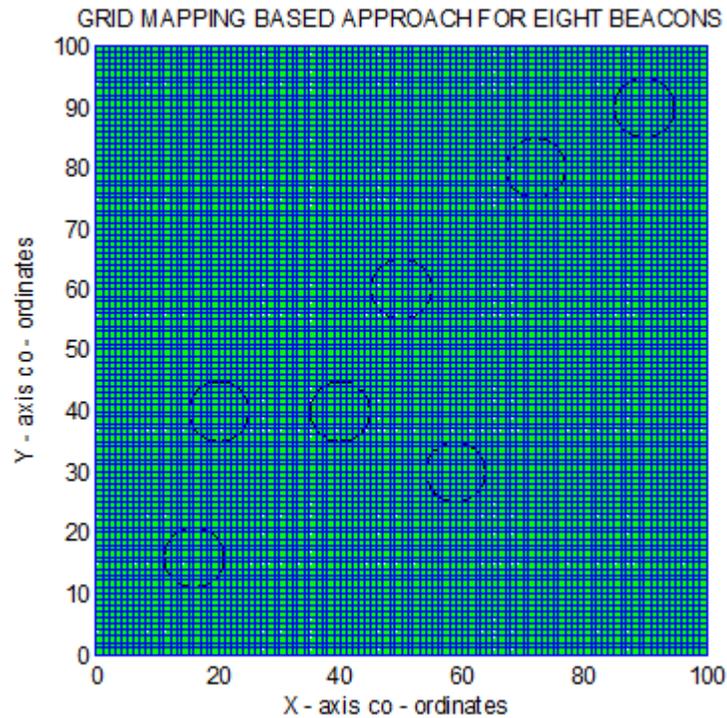

Figure 1. An example of Grid mapping based approach for eight beacons.

## 3. SIMULATIONS AND RESULTS

To understand the behavior of uncertainty, graphical analysis is done with MATLAB Version 7.6.0.324 (R2008a). Extensive simulations and numerical experiments in the order of 500s are carried out with random number of beacons starting from least number varying in steps and also increasing beacon radius. The function of localization uncertainty $f(r, b)$, is calculated by using two parameters one is beacon radius, $r$ and other one is number of beacons, $b$. Then graph is plotted with varying number of beacons v/s uncertainty in localization and also varying beacon radius v/s uncertainty in localization to have a clear picture about the least possible uncertainty that can be obtained in localization. This analysis gives the optimum beacon radius and optimum number of beacons required to achieve the least possible localization uncertainty. Standard deviation in uncertainty is calculated to know the variation in localization uncertainty. And COV is also calculated.

Signal & Image Processing : An International Journal (SIPIJ) Vol.6, No.3, June 2015

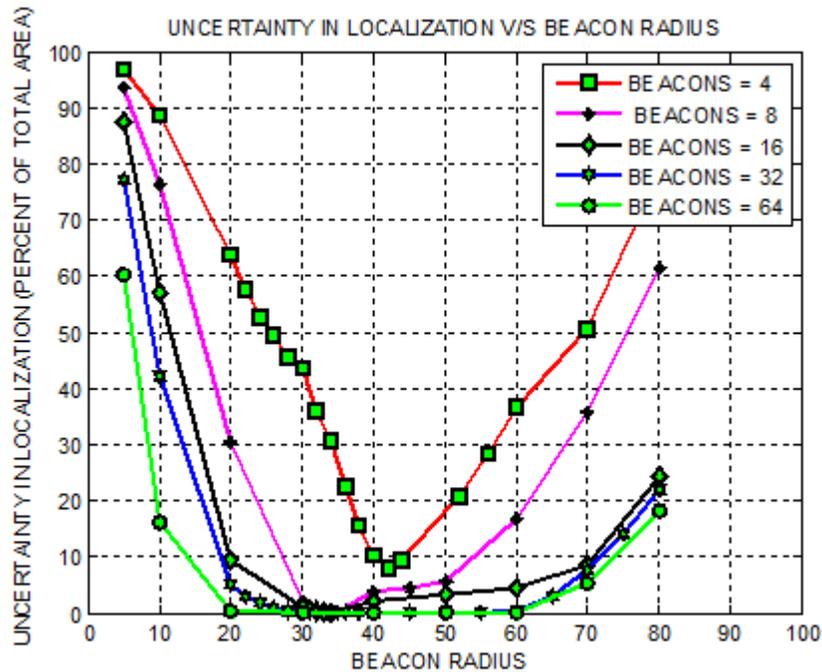

Figure 2. Uncertainty in localization v/s beacon radius

Figure 2 shows the graph of localization uncertainty verses beacon radius (normalized). As can be seen from the Figure 2, uncertainty in localization will be very high at very small beacon radius and it goes on decreasing when the beacon radius increases, but at a particular beacon radius uncertainty will be very less, this is the least possible uncertainty that can be achieved and the corresponding radius is the optimum beacon radius and after that the uncertainty again increases with increase in beacon radius. For instance if number of beacons = 4, then localization uncertainty is calculated for different radius starting from least one, say 5 (from Figure 2). This experimentation is done for different number of beacons like 8, 16, 32, 64, and so on.

To understand some critical regions of the graph shown in Figure 2, all the curves of Figure 2 are plotted individually in a particular beacon radius range where analysis is required i.e., to have optimum beacon radius. Figure 3 shows a graph of uncertainty in localization verses beacon radius for number of beacons = 4, where the optimum beacon radius would be 42, which gives the least possible uncertainty. Similarly, Figure 4 is for number of beacons = 8, where the optimum beacon radius would be 35. Figure 5 talks on number of beacons = 16, where the optimum beacon radius would be 33. Figure 6 shows for number of beacons = 32, where the optimum beacon radius would be ranging from 30 to 40. Finally Figure 7 gives a view on number of beacons = 64, where the optimum beacon radius would be ranging from 20 to 30, which gives the least possible uncertainty in localization.

16Signal & Image Processing : An International Journal (SIPIJ) Vol.6, No.3, June 2015

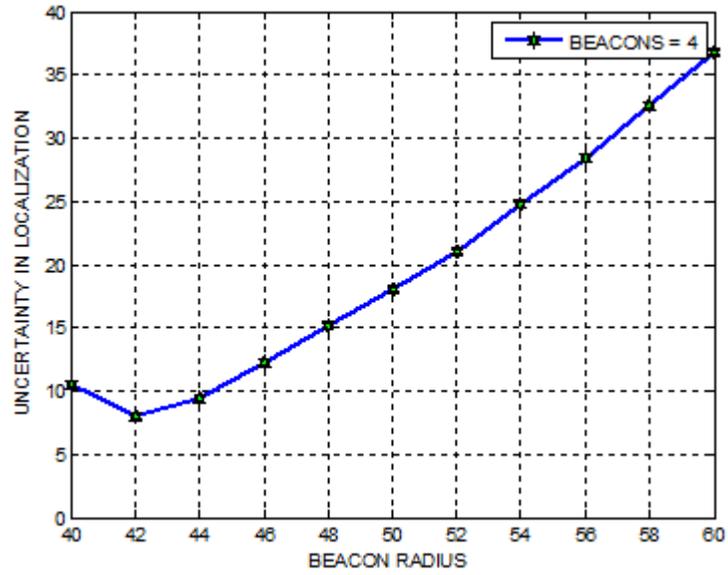

Figure 3. Uncertainty in localization v/s beacon radius for number of beacons = 4

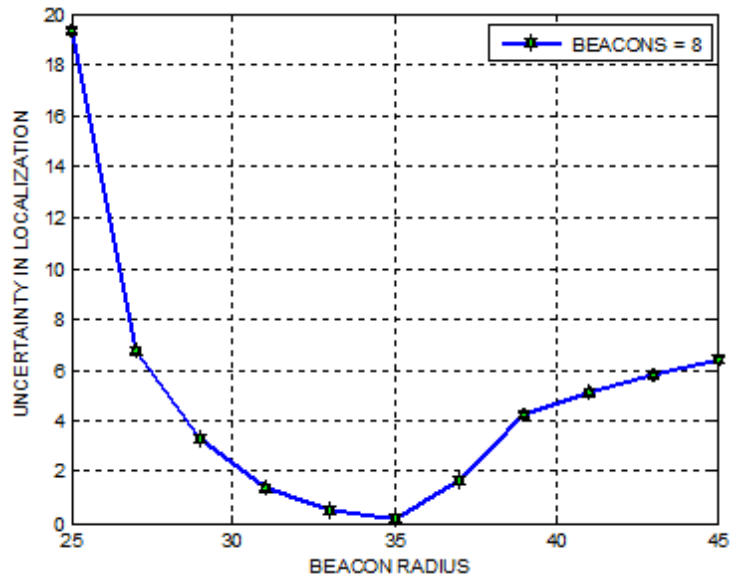

Figure 4. Uncertainty in localization v/s beacon radius for number of beacons = 8

Signal & Image Processing : An International Journal (SIPIJ) Vol.6, No.3, June 2015

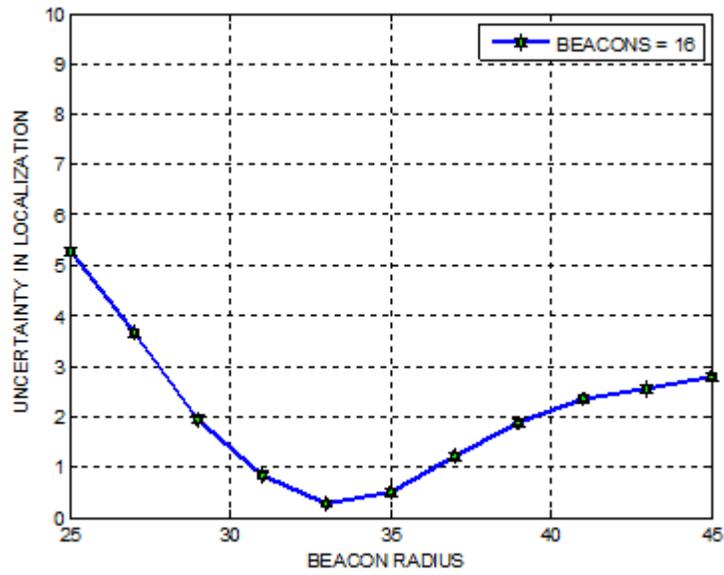

Figure 5. Uncertainty in localization v/s beacon radius for number of beacons = 16

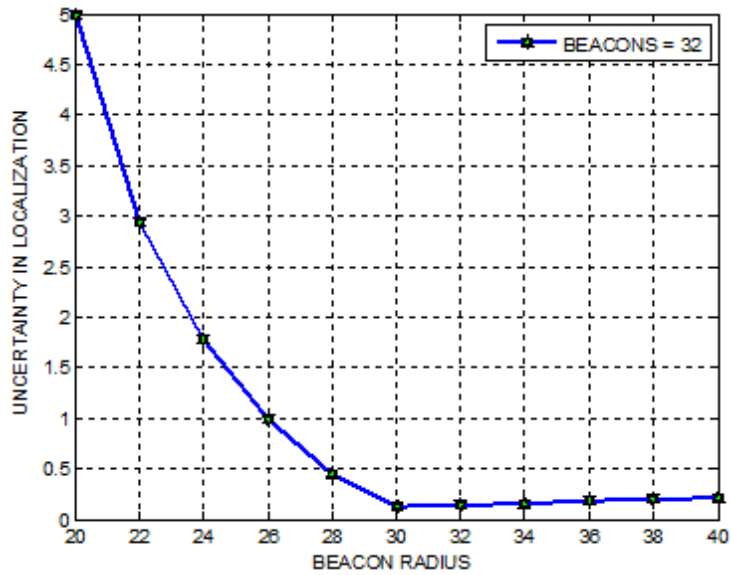

Figure 6. Uncertainty in localization v/s beacon radius for number of beacons = 32

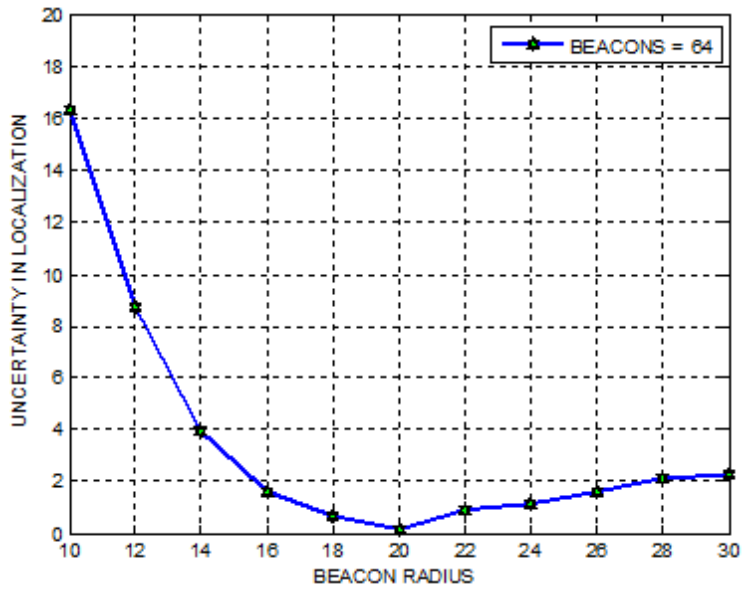

Figure 7. Uncertainty in localization v/s beacon radius for number of beacons = 64

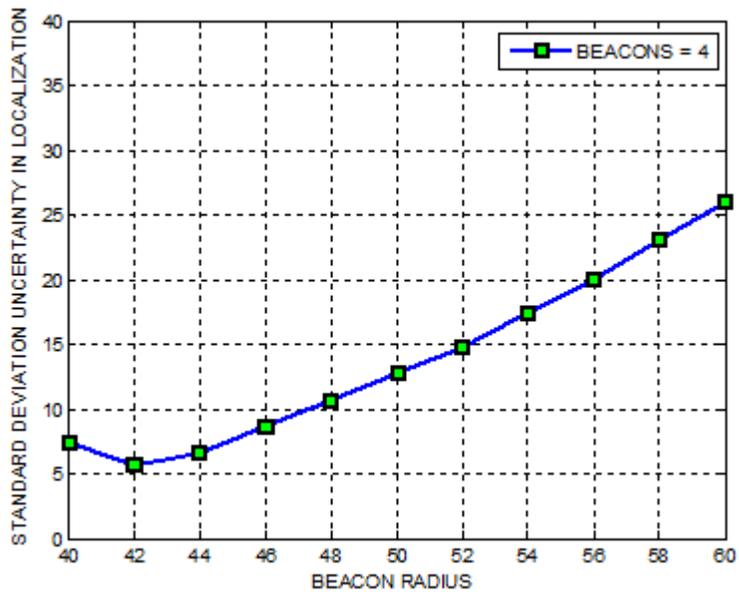

Figure 8. Standard deviation uncertainty in localization v/s beacon radius for number of beacons = 4

Figure 8 shows a graph of standard deviation in uncertainty verses beacon radius for number of beacons = 4. Comparing Figure 3 with Figure 8, we can notice that the deviation in percentage uncertainty occurs for a particular beacon radius. For instance, for r = 40 and b = 4 (from Figure 3 and Figure 8) the deviation in percentage uncertainty is from 7.4034 to 10.4700. Another example is, for r = 60 and b = 4 (from Figure 3 and Figure 8) the percentage uncertainty varies from 26.0145 to 36.7900. If r = 42, the localization uncertainty in percentage varies from 8.0400 to 5.6851.

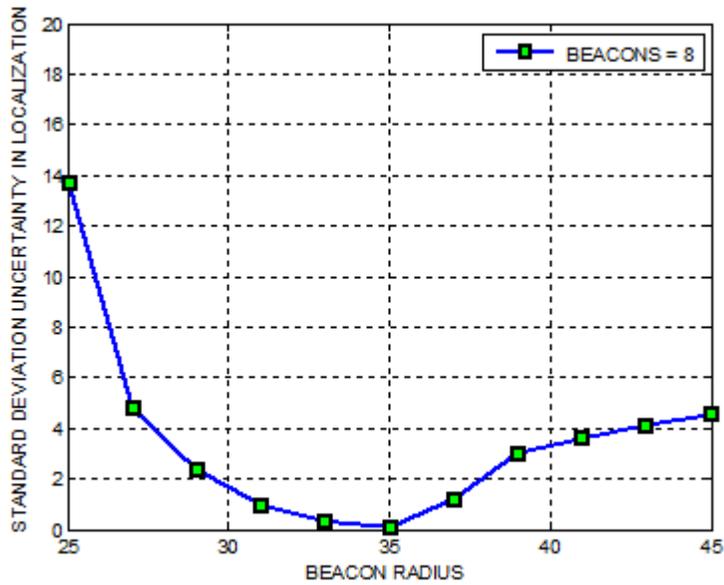

Figure 9. Standard deviation uncertainty in localization v/s beacon radius for number of beacons = 8

Comparing Figure 4 with Figure 9, for r = 25 and b = 8 the percentage uncertainty varies from 13.6684 to 19.3300. Another example is, for r = 45 and b = 8 (from Figure 4 and Figure 9) the percentage uncertainty varies from 4.5255 to 6.4000. If r = 35, the localization uncertainty in percentage varies from 0.1200 to 0.0849.

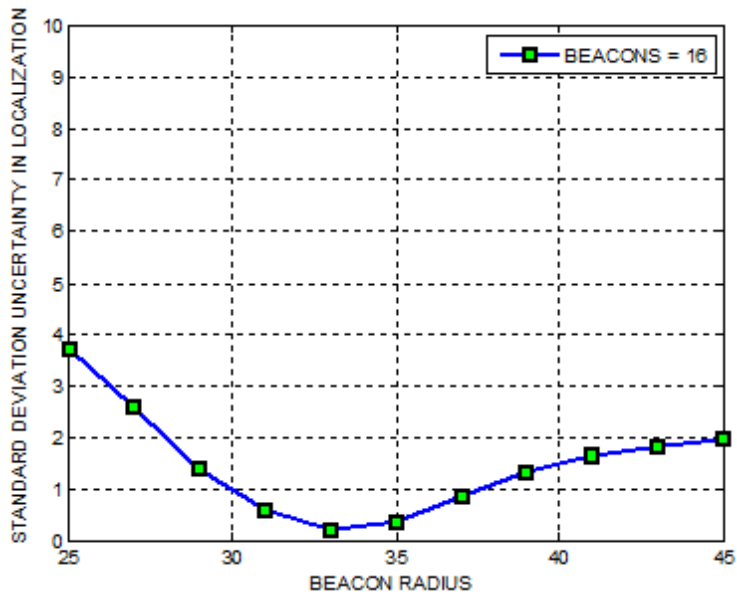

Figure 10. Standard deviation uncertainty in localization v/s beacon radius for number of beacons = 16

Comparing Figure 5 with Figure 10, for r = 25 and b = 16 the percentage uncertainty varies from 3.7335 to 5.2800. Another example is, for r = 45 and b = 16 (from Figure 5 and Figure 10) the percentage uncertainty varies from 1.9658 to 2.7800. If r = 33, the localization uncertainty in percentage varies from 0.2900 to 0.2051.

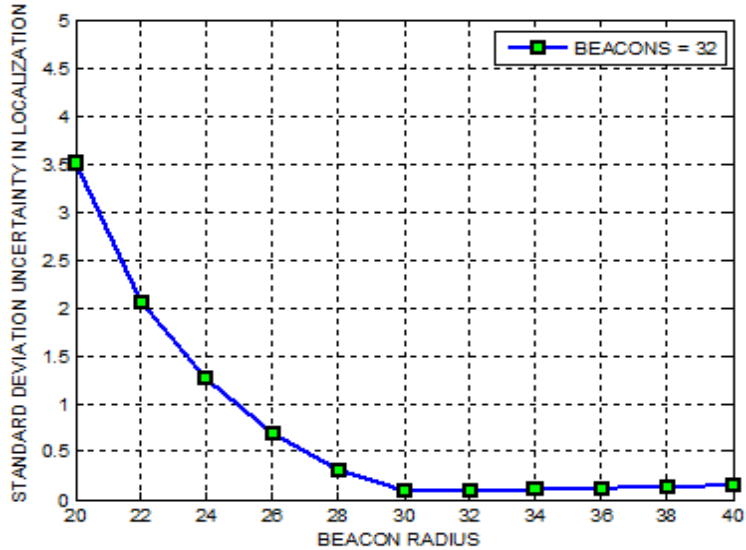

Figure 11. Standard deviation uncertainty in localization v/s beacon radius for number of beacons = 32

Comparing Figure 6 with Figure 11, for r = 20 and b = 32 the uncertainty varies from 3.5214 to 4.9800. Another example is, for r = 40 and b = 32 (from Figure 6 and Figure 11) the uncertainty varies from 0.1485 to 0.2100. If r varies from 30 to 40, the localization uncertainty in percentage varies from 0.0919 to 0.1485 and 0.1300 to 0.2100 respectively.

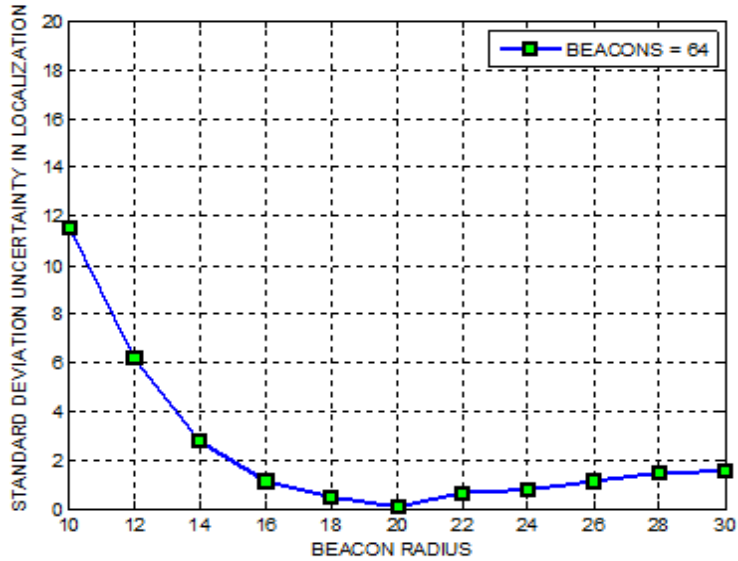

Figure 12. Standard deviation uncertainty in localization v/s beacon radius for number of beacons = 64

Comparing Figure 7 with Figure 12, for r = 10 and b = 64 the uncertainty varies from 11.5612 to 16.3500. Another example is, for r = 30 and b = 64 (from Figure 7 and Figure 12) the uncertainty varies from 1.5768 to 2.2300. If r varies from 20 to 30, the localization uncertainty in percentage varies from 0.1500 to 2.2300 and 0.1061 to 1.5768 respectively.

Table 1 summarizes the comparison of optimum radius and least possible percentage uncertainty that can be achieved and standard deviation in percentage uncertainty for different ranges of beacon radius and number of beacons.

Table 1.Comparison of optimum radius and %U for different ranges of beacon radius and number of beacons

| b | r | Optimum r | % U | STD %U |
|---|---|---|---|---|
| 4 | 40 – 60 | 42 | 8.0400 | 5.6851 |
| 8 | 25 – 45 | 35 | 0.1200 | 0.0849 |
| 16 | 25 – 45 | 33 | 0.2900 | 0.2051 |
| 32 | 20 – 40 | 30 – 40 | 0.1300 – 0.2100 | 0.0919 – 0.1485 |
| 64 | 10 – 30 | 20 – 30 | 0.1500 – 2.2300 | 0.1061 – 1.5768 |

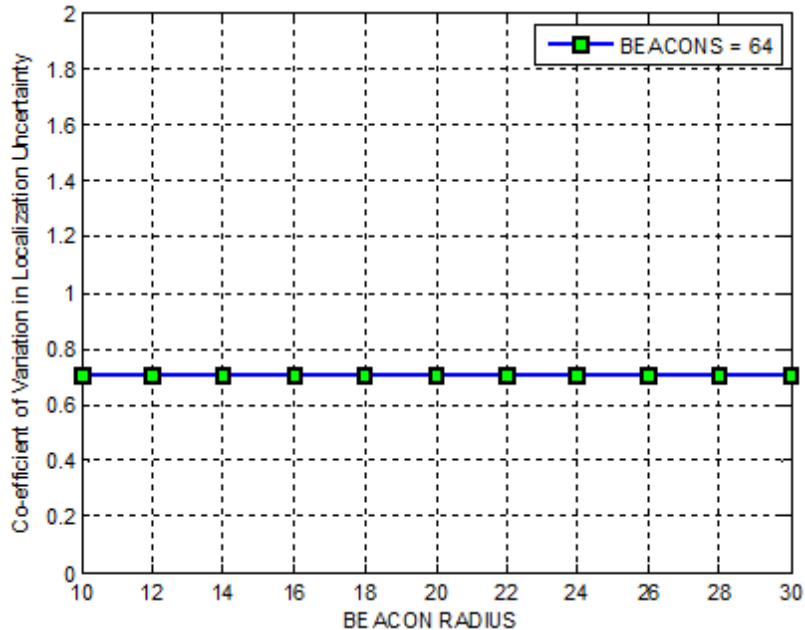

Figure 13. Co-efficient of variation in localization uncertainty v/s beacon radius for number of beacons = 64

The Co-efficient Of Variation (COV) is defined as the ratio of the standard deviation to localization uncertainty. COV is a dimensionless number. Figure 13, shows the graph of COV

Signal & Image Processing : An International Journal (SIPIJ) Vol.6, No.3, June 2015

verses beacon radius for number of beacons = 64. The value of COV here is 0.7071, not only for 64 beacons, for 4, 8, 16, 32 beacons also the COV remains the same.

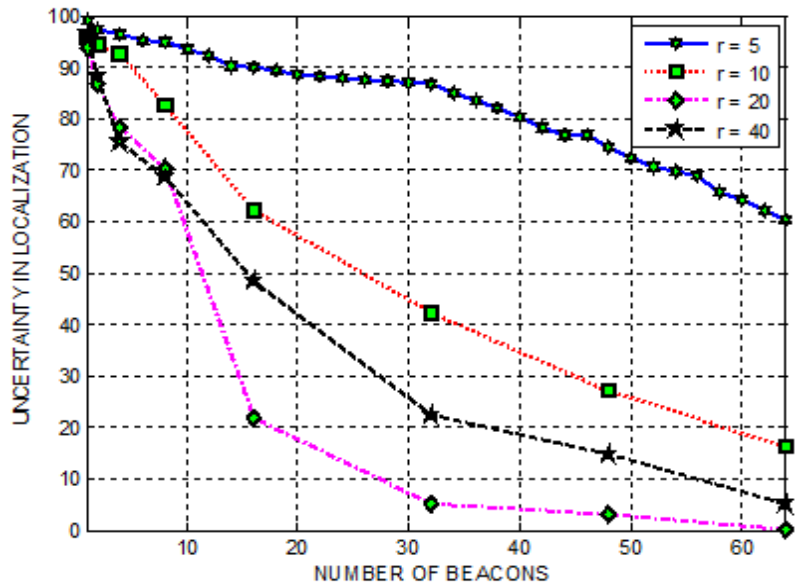

Figure 14. Uncertainty in localization v/s number of beacons

Figure 14 shows a graph of localization uncertainty verses number of beacons for beacon radii of 5, 10, 20 and 40. From Figure 14, initially at less number of beacons the uncertainty is very large and it goes on decreasing with increasing number of beacons. Here the curves may cross each other due to intersection or overlap of random beacons that leads to having same uncertainty value.

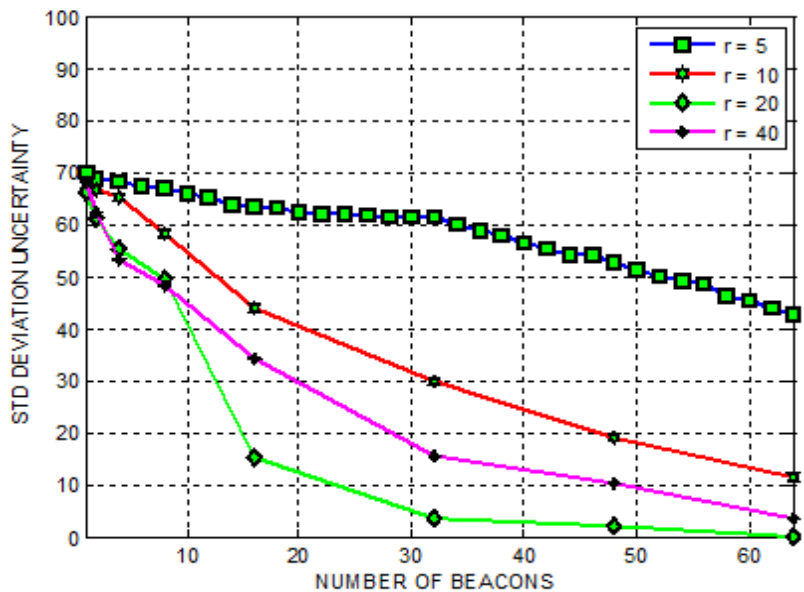

Figure 15. Standard deviation uncertainty in localization v/s number of beacons

23

Figure 15 shows a graph of standard deviation in uncertainty verses number of beacons. Comparing Figure 14 and Figure 15, we can notice that how the deviation in uncertainty has occurred.

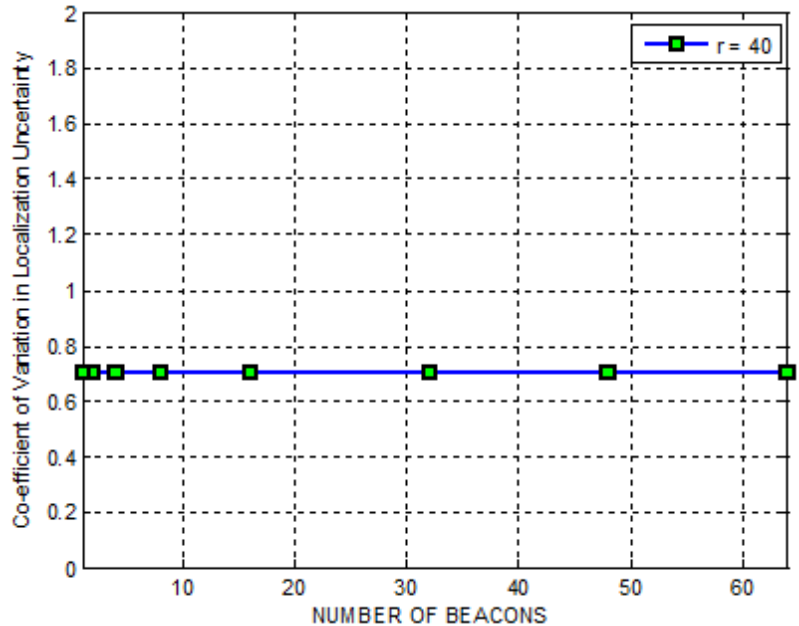

Figure 16. Co-efficient of variation in localization uncertainty v/s number of beacons for r = 40

Figure 16, shows the graph of COV verses number of beacons for beacon radius = 40. The value of COV here, it is also 0.7071 i.e., for beacon radius = 40. For r = 5, 10, 20 also the COV remains the same.

## 4. CONCLUSION

Uncertainty in localizing a target is a problem in Wireless Sensor Networks which is very difficult to handle. Severe uncertainties cause disasters in Wireless Sensor Networks. Localizing a target with less uncertainty from known beacon locations is a challenging problem. Care must be taken while handling uncertainties and here we have designed a simple model and achieved least possible uncertainty by choosing optimum beacon radius and also optimum number of beacons by conducting extensive simulations and numerical experiments. Deviations in uncertainty are also discussed and validated followed by Co-efficient Of Variation. Extending this work to build a rigid and hard enough resistive model so that RSS and uncertainty in predicting the target should not vary with changes in environmental conditions like temperature, pressure, humidity, rainfall, fog, mist and natural calamities like earthquake, volcanic eruption and many more. Over and all signal processing or target localization in wireless sensor networks places an important role in today's science and technology.

**AUTHORS**

Santhosh N Bharadwaj completed his bachelor degree in Electronics and Communication engineering from Visvesvaraya Technological University, Belgaum, Karnataka in the year 2012. He is persuing his Master of Technology in Network and Internet Engineering. His area of interests includes Signal Processing and Wireless Sensor Networks.

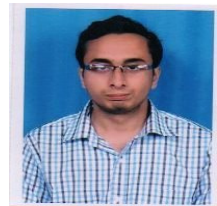

Dr. S. N. Jagadeesha received his B.E., in Electronics and Communication Engineering, from University B. D. T College of Engineering., Davangere affiliated to Mysore University, Karnataka, India in 1979, M.E. from Indian Institute of Science (IISC), Bangalore, India specializing in Electrical Communication Engineering., in 1987 and Ph.D. in Electronics and Computer Engineering., from University of Roorkee (I.I.T, Roorkee), Roorkee, India in 1996. He is an IEEE member and Fellow, IETE. His research interest includes Array Signal Processing, Wireless Sensor Networks and Mobile Communications. He has published and presented many papers on Adaptive Array Signal Processing and Direction-of-Arrival estimation. Currently he is Professor in the Department of Computer Science and Engineering, Jawaharlal Nehru National College of Engg. (Affiliated to Visvesvaraya Technological University), Shimoga, Karnataka, India.

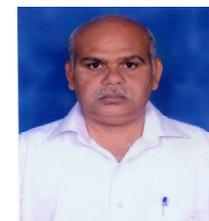

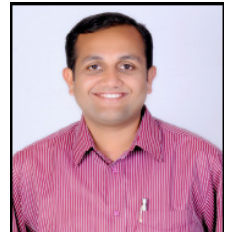

Ravindra. S. received his B.E., in Electrical and Electronics Engineering., and M.Tech., in Networking and Internet Engineering, from Visvesvaraya Technological University, Belgaum, Karnataka, India in 2006 and 2008 respectively. He is currently working towards his Doctoral Degree from Visvesvaraya Technological University, Belgaum, Karnataka, India. At present he is working as Assistant Professor, in Computer Science and Engineering department of Jawaharlal Nehru National College of Engineering  (affiliated to Visvesvaraya Technological University), Shimoga, Karnataka, India.